\documentclass[useAMS,usenatbib, usegraphicx]{mn2e}

\usepackage{amsmath}
\usepackage{graphicx}
\usepackage{xfrac}
\usepackage{dblfloatfix}
\usepackage{lastpage}
\usepackage{placeins}
\title[The Observational Effects and Signatures of Tidally Distorted Solid Exoplanets]{The Observational Effects and Signatures of Tidally Distorted Solid Exoplanets}
\author[Prabal Saxena, Peter Panka and Michael Summers]{Prabal Saxena\thanks{E-mail:
psaxena2@gmu.edu}, Peter Panka and Michael Summers\\
School of Physics, Astronomy and Computational Science, George Mason University, Fairfax, VA, 22030, USA}
\begin{document}

\pagerange{\pageref{firstpage}--\pageref{lastpage}} \pubyear{2014}

\maketitle

\label{firstpage}

\begin{abstract}
Our work examines the detectability of tidally distorted solid exoplanets in synchronous rotation.  Previous work has shown that tidally distorted shapes of close-in gas giants can give rise to radius underestimates and subsequently density overestimates for those planets.  We examine the assumption that such an effect is too minimal for rocky exoplanets and find that for smaller M Class stars there may be an observationally significant tidal distortion effect at very close-in orbits.  We quantify the effect for different stellar types and planetary properties using some basic assumptions.  Finally, we develop a simple analytic expression to test if there are detectable bulge signatures in the photometry of a system.  We find that close in for smaller M Class stars there may be an observationally significant signature that may manifest itself in both in-transit bulge signatures and  ellipsoidal variations. 
\end{abstract}


\section{Introduction}

The distortion of an astronomical object's spherical shape has long been known to provide insight to the interior structure, composition and evolution of a body.  Research exploring shape has spanned over many different types of objects including gaseous bodies \citep{cha33, lai93} as well as solid, solar system bodies \citep{der79}.  Interest in the shape of an object as a proxy for other parameters has extended to the burgeoning field of exoplanets.

Theoretical and modeling work has already yielded a number of predictions and constraints on the ability to detect rotational and tidal deformation in gas giant exoplanets. Additionally, similar work has also explored the potential effects of such distortions on the ability to accurately describe an exoplanet system.  Early estimates of a detectable  variation in transit depth within a light curve due to oblateness yielded results which even for the most favorable scenarios were very close to observational limits \citep{sea02, bar03}.  On the other hand, the reflection effect for tidally distorted planets has been shown to be potentially significant in certain cases \citep{bud11}.  In that study, Budaj demonstrated the significant light curve variations that arise for Roche Model approximated tidally distorted planets when the bond albedo of the planet is varied to simulate different levels of reflected starlight versus absorbed/re-radiated/redistributed light.  Indeed, there have been observational results which have suggested potential detection of just such a deformation in the shape of a gas giant exoplanet.  Ellipsoidal variations in transit depth of the light curve of WASP-12b \citep{cow12} have been roughly fitted and suggest a near 3:2 ratio between the planet's longest and shortest axes.

In addition to these variations, a number of studies concluded that the gravitational quadrupole field which would arise from tidal and rotational bulges could induce apsidal precession that would manifest itself as predictable transit timing effects \citep{jor08, rag09, kan12}.  In addition to apsidal precession, exoplanet oblateness and obliquity would also induce spin precession that under certain conditions could yield detectable signals for certain gas giants \citep{car101}.  Observational results have been used to constrain the oblateness of HD189733b through non-detections of transit depth changes due to such precession and by non-detection of potential transit timing variations \citep{car102}.

An important effect of potential distortions to the shape of an exoplanet is the potential for particular orbital orientations to bias the volume measurement of a planet and subsequently other derived parameters.  For planets in a synchronous orbit that are close in to the host star the projected area of the ellipsoidal shape of the planet caused by rotational and tidal effects can result in an underestimate of the radius of the planet \citep{lec11}.  In some cases these distortions may result in biases of 10\% or higher in the light curves and radius measurements of these planets \citep{li10}.  Parameters which are subsequently derived from these radius estimates such as density and size of the solid inner cores of gas giants can then also be estimated incorrectly \citep{bur14}.  While such an effect has been explored for gas giants, one of the main topics of this study is to look at whether this bias may also exist for solid exoplanets with distorted shapes.  In addition we also examine the assumption that the effects of an aspherical solid exoplanet in its transit photometry are too small to be observationally significant.  In particular, we attempt to quantify the variation in the projected area of a solid planet within its light curve.

Solid exoplanets constitute a significant portion and increasing portion of the current census of planets outside our solar system. Characterization of a planet as a likely solid planet is obtained using bulk properties of a planet such as density from observed constraints on radius and mass \citep{val07}.  While degeneracy over interior models still remain, such a parameterization can to first order divide a planet between a gaseous or solid planet.  Estimates of occurrence rates for planets vary significantly based on the spectral class of the host star and the orbital separation of the planet from the star \citep{dre13, mul14}.  However, bound planets have been detected at distances from almost right at the Roche limit for their host star \citep{hel11, gil14, rap13} to much farther out from their host star than any planet in our solar system.  This substantial parameter space which we are faced with when trying explore the potential bias and observable signatures of distorted aspherical solid planets is resolved by the fact that previous work examining shape distortions on gas giant planets have required certain orbital and observational constraints which also neatly align with the bias of transit photometry to detect close-in exoplanets.

In section 2 of this article we explore observational bias effects in the measured radius of solid exoplanets due to planetary asphericity.  We address the theory and assumptions made regarding the orbits, rotation and shape distortions of these planets.  We then produce results quantifying observational bias and subsequent bias in estimated planetary parameters for a range of different sized exoplanets around different spectral classes.  In section 3 we address whether the possibility and magnitude of any in-transit signal that can be detected due to these distortions to planetary shape and discuss the potential for ellipsoidal variations.  Finally, in section 4 we comment on the results of our estimates and discuss implications and potential future of areas of study.

\section{Measuring Observational Bias in Radius Estimates}
\subsection{Tidal and Rotational Deformation theory and assumptions}

A rotating planet orbiting a star will be subject to both tidal and rotational forces which distort the shape of the planet.  Since the gravitational potential due to the star varies inversely with distance, the resulting gradient across the discrete boundaries of the orbiting planet will induce a symmetric tidal bulge in the direction of the tide-inducing body that will deform the object into a prolate ellipsoid.  The gravitational potential V felt by a point P on the surface of a planet due to some external body treated as a point mass such as the host star can be written as

\begin{equation}
V = -G \dfrac{m_{s}}{a} [1 + \dfrac{R_{p}}{a} \cos\psi + (\dfrac{R_{p}}{a})^{2}\dfrac{1}{2}(3\cos^{2}\psi-1)+...] 
\end{equation} \begin{center} \citep{mur99} \end{center}

Where G is the gravitational constant, $m_{s}$ is the mass of the star, a is the orbital separation of the two objects, $R_{p}$ is the radius of the planet, and $\psi$ is an angle measured from the line joining the centers of the two bodies to the point on the surface.  

The potential here is the result of binomially expanding the distance from the perturbing body to the point on the surface, which assumes $\sfrac{R_{p}}{a} \ll{1}$.  This will be a generally well held assumption for nearly all the cases we examine (for example a 2 Earth Radii planet orbiting at the fluid Roche Limit of a M1 V star will have a $\sfrac{R_{p}}{a}$ approximately equal to that of the Earth-moon system).  Even for the scenario with largest and closest orbiting planet around the smallest host star, the value of $\sfrac{R_{p}}{a}$ is $<0.1$.
The tide raising part of the potential is the $2^{nd}$ order term and can be written in terms of the Legendre polynomial of degree two given by:

\begin{equation}P_{2}(\cos\psi) = \dfrac{1}{4}(3\cos2\psi +1)\end{equation}

In order for the surface of the planet to be on an equipotential the deformed shape of the planet can then be described by the same solid spherical harmonic of the second degree that describes the tide raising potential.  Taking into account implied symmetry about the line joining the two bodies from the dependence on the angle $\psi$ we then describe the deformed shape of an incompressible planet as

\begin{equation}R_{cb}(\psi) = A[1+S_{2}P_{2}(\cos\psi)]\end{equation}

\begin{figure*}
  \begin{center}
      \includegraphics[angle=90,scale=0.45]{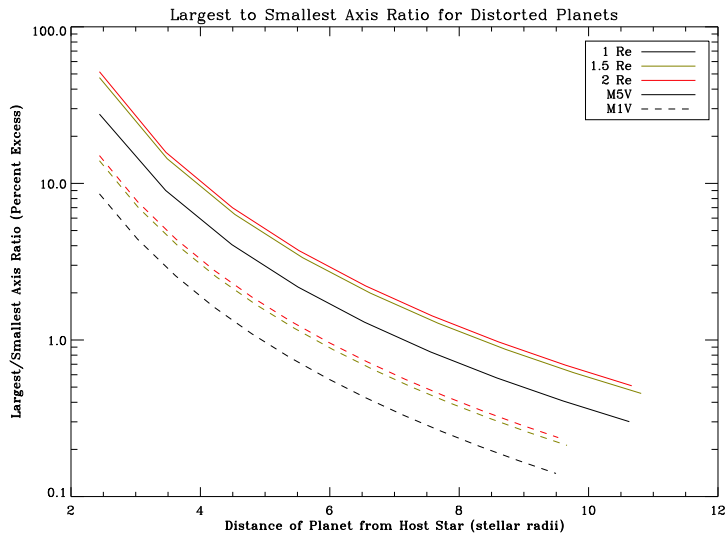}
      \caption{The ratio of the longest to smallest axis for triaxial ellipsoid planets of 1, 1.5 and 2 Earth Radii orbiting M5V and M1V stars. For 1.5 and 2 Earth Radii planets orbiting an M5V star the ratio can approach 3:2 near the fluid roche limit.  Around M1V stars a longest axis about 10\% larger than the shortest axis is typical for all three planets near the fluid roche limit.}
     \label{fig:axisratios}
  \end{center}
\end{figure*}

Where R is the distance of the surface from the center of the planet, A is the mean radius and $S_{2}$ is a constant which describes the ellipticity of the planet's shape.  \citet{mur99} also extend the treatment of the planet as a two layer incompressible body.  They take into account additional forces acting at the boundary of the two layers including variational forces within the layers leading to loading terms as well as the internal potentials of each layer with their own separate densities.  For our study we use a first order calculation of the deformed shape of the body as a one layer uniform body – equivalent to the scenario described by Murray and Dermott of the deformation of the core boundary if the ocean were removed.  Due to the degeneracy of potential interior structures of exoplanets when just given bulk values such as mass and volume we find it sufficient to characterize the shape deforming factors with as few variables as possible.

In the case of a planet modeled with a single uniform layer the equilibrium tide and consequently the ellipticity of the planet can be characterized by just the orbital separation of the bodies, bulk parameters of mass and mean radius and the effective rigidity of the planet.  The amplitude of the equilibrium tide induced by the host star is just 

\begin{equation}AS_{2} = \dfrac{(5/2)\zeta _{c}}{1+\tilde{\mu}}\end{equation}

where

\begin{equation}\zeta = \dfrac{m_{s}}{m_{p}} (\dfrac{R_{p}}{a})^{3}R_{p}\end{equation}

Where $m_{p}$ is the mass of the planet and the dimensionless variable $\tilde{\mu}$ is the effective rigidity, which relates the elastic strength of the planet to its gravity and which is given by:

\begin{equation}\tilde{\mu} = \dfrac{19\mu}{2\rho g_{c}A}\end{equation}

where $\rho$ is just density of the planet, g is just the planet's gravitational constant, and $\mu$ is the rigidity.  We assume a value of $\mu$ of $\sim100$ GPa, about the mean value for the Earth \citep{fow05}, for the planets we test.  We choose this value in order to make as few assumptions about the compositional and structural nature of these planets given the current inability to observationally constrain those parameters (of course choosing an Earth-like rigidity value is an assumption in itself). Additionally, given that current hypotheses regarding the nature of these planets range from ostensibly rigid Iron-core Hot Jupiter remnants to planets which may significantly be less rigid due to partial or wholescale melting any assumed rigidity value would be somewhat arbitrary.  We do however also test a slightly higher value of $\mu$ $\sim300$ GPa, the predicted rigidity value of the Earth's core \citep{bul69, gol68}, in order to test the effect of different rigidity values in terms of the change in bias of observed planetary parameters and whether direct observations of elliptical shape would be sensitive enough to constrain mean planetary rigidity values.

\begin{figure*}
  \begin{center}
      \includegraphics[angle=90,scale=0.55]{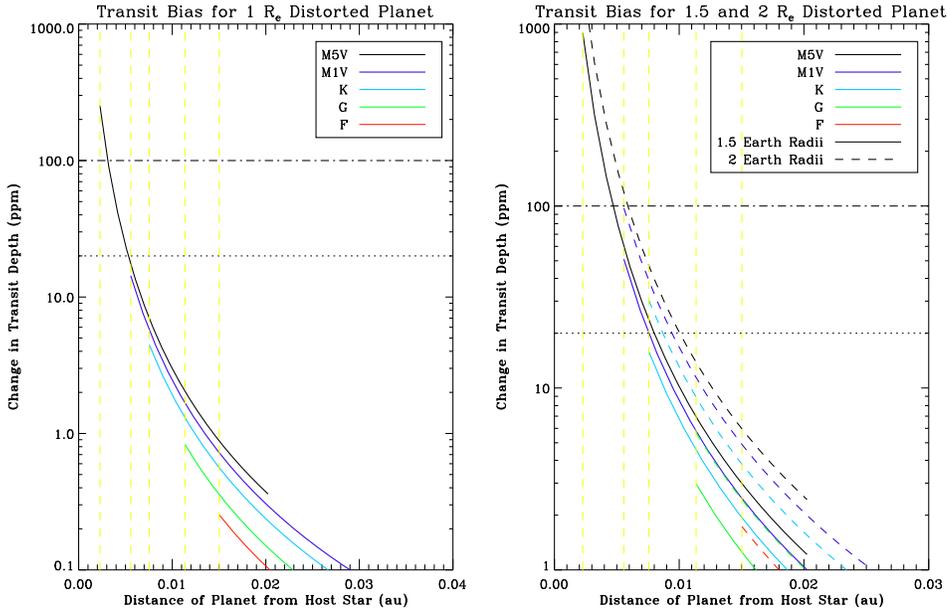}
      \caption{Transit biases for 1, 1.5 and 2.0 Earth Radii planets due to shape distortion from tidal and rotational effects.  The dotted and dash dotted lines are 20 and 100 ppm levels.  The vertical dashed yellow lines indicate the fluid Roche Limit for an M5V, M1V, K, G and F class star going from left to right for Earth like densities.  Due to the one third power dependence on ratio of densities, the fluid Roche Limit for 1.5 and 2 Earth Radii planets isn't significantly different.  The right panel contains transit bias values for 1.5 and 2 Earth radii}
     \label{fig:difftransitdepth}
  \end{center}
\end{figure*}

Once we have found the constants quantifying the tidal effect on planetary shape we can now use an additional assumptions of synchronous rotation of these planets in a near-equatorial (with zero impact parameter), near circular orbit around their host stars and easily add rotational deformation contributions to the planet's shape.  It can be shown that the rotational potential responsible for deformation at a point within the planet can also be written as a second degree solid harmonic \citep{der79}.  In fact the only difference between the two potentials is that the rotational deformation is 1/3 the magnitude of the tidal potential and has a different axis of symmetry (in the case we refer to here it is the z axis).  As a result we can use the same shape function of the planet as we did before and linearly add the rotational and tidal contributions for the corresponding angles along their separate axes of symmetry.  The resulting triaxial ellipsoid then has its axes parameterized as follows:

\begin{equation}a = A(1 + \dfrac{7S_{2}}{6})\end{equation}	
\begin{equation}b = A(1 - \dfrac{S_{2}}{3})\end{equation}
\begin{equation}c = A(1 - \dfrac{5S_{2}}{6})\end{equation}

where a is the axis pointed towards the host star, b is the axis orthogonal to it in the planet-star system's equatorial plane and c is the orthogonal axis of rotation of the planet. 

The synchronous orbit assumption ensures a special geometric configuration where the longest axis of the planet is always oriented towards the star - this ensures the mid transit alignment from the perspective of the observer will yield a predictable projected area of the planet transiting its host star given by the product of the two other orthogonal axes.   The bias in transit depth at mid transit for a planet with mean radius $R_{p}$ is then just given by

\begin{equation}\dfrac{(bc)-(R_{p}^{2})}{R_{s}^{2}}\end{equation}

where $R_{s}$ is the mean radius of the star.  This bias was calculated for a number of different planet sizes and different spectral classes.  In particular, bias values were calculated for 1, 1.1447, 1.5, 2, 2.154 and 2.25 $R_{e}$ solid planets with bulk densities taken from the relation given in equation 1 of \citet{wei14}.  While that density relation was for putative terrestrial planets smaller then 1.5 $R_{e}$, we extend it to larger planets since the question of whether large planets close in can retain a significant gaseous envelope is still open.  These bias values were calculated for 6 different main sequence spectral classes: M6, M5, M1, F, G, and K.   Most of the images in this study will output results for the second smallest star (M5 V) as confirmed planet detections have only been made for stars down to that spectral class and because that is the limiting spectral type the \textit{Transiting Exoplanet Survey Satellite (TESS)} \citep{ric14}  will be sensitive to.  Main sequence stars larger than K class stars were not tested as any bias or distortion signal would be too minimal given the large projected area of the stars - an assumption that is well founded based upon the following results and currently observed upper bounds on radii of solid planets \citep{dum14}.  Physical parameters for M dwarfs were taken from literature \citep{rei05} while the parameters used for the F, G and K class stars were from well-known examples (61 Cygni (K5 V), The Sun (G2 V), Pi3 Orionis (F6 V) respectively).  Finally, orbital distances for which tidal bulges and biases were calculated were based on an inner distance defined by the fluid Roche limit of each star-planet system and an outer distance at which a 1.5 $M_{e}$ planet with the corresponding fitted bulk density would be tidally locked within 1 billion years around different stars \citep{gla96}. An initial spin rate of one rotation per 13.5hrs was used \citep{gol66} based on the predicted initial spin state for Venus.  This is a highly speculative choice but was made because the types of planets examined in this study are much larger than Mercury and because the most appropriate analogue choice, the Earth, has had its rotational history significantly modified by interactions with the moon.  In any event, initial rotation rates were varied from 1.5 hrs to 1 day and found to have minimal effect on the outer radius limit tested due to the one sixth power they are raised to.  In addition, the induced tidal bulges raised closer to the outer distances are minimal for the purpose of this study.  The fluid Roche limit is a conservative assumption as solid planets have been found interior to it \citep{rap13} while bias and bulge estimates are minimal for the outer distance.

The importance of the synchronous rotation assumption necessitates a discussion of the likelihood of such a state.  Recent work has shown that pseudosynchronous rotation for terrestrial bodies is an unphysical state based on an improper treatment of the tidal torque \citep{mak131} and that most planets with a high enough eccentricity fall into higher order resonances.  Given that many close-in planets have potentially been excited to higher eccentricities during their orbital history before their orbits were circularized this may mean that even for close-in planets synchronous rotation may be an atypical state. 

\begin{figure*}
  \begin{center}
      \includegraphics[angle=90,scale=0.55]{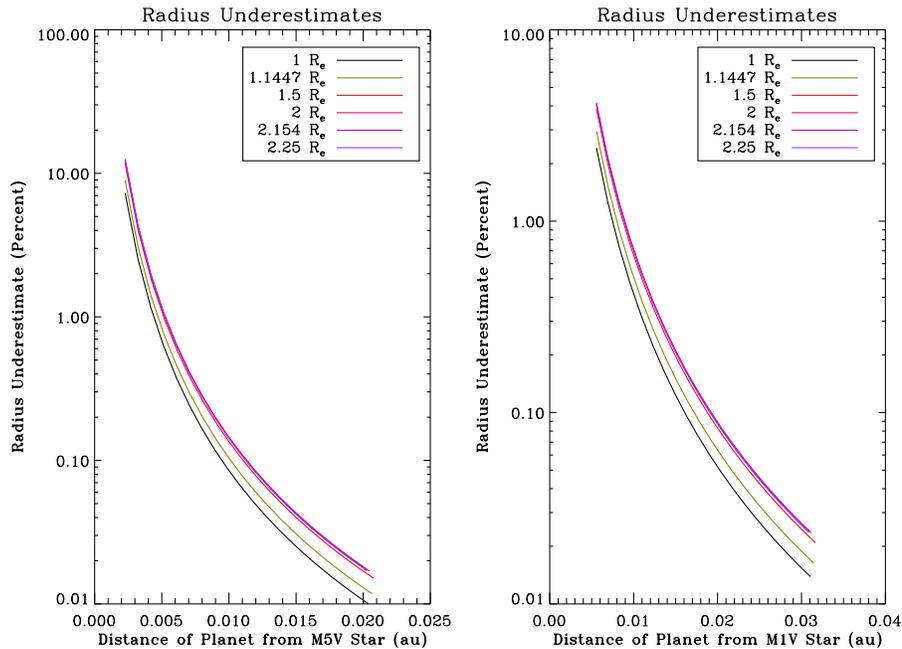}
      \caption{Radius underestimates for different sized solid planets due to tidal and rotational distortions. The left image is for synchronously rotating planets around M5 V stars and right for M1 V stars.}
     \label{fig:radiusunder}
  \end{center}
\end{figure*}

However, there are different scenarios under which such planets which may fall into synchronous rotation - while these are used as justification for such an assumption, the potential to detect bulge signatures may also act as a test for the spin state.  In particular, planets on very circular orbits with low eccentricity, planets that at some point were thrown into retrograde orbit (either by migration or capture), and bodies that were relatively cold and not very responsive to tidal torques may have fallen into synchronous rotation. Importantly, significantly triaxial bodies may also have a triaxiality caused torque if their average inertia axis is tilted with respect to the primary-secondary axis that can then compensate for the nonzero secular tidal torque and result in a synchronous orbit (Efroimsky 2014, private comm.).

\subsection{Bias Results}

Estimates of the distortion to a planet's shape around M5V and M1V stars are given in figure \ref{fig:axisratios}.  These are the cases where the shape of the planet is the most deformed and where the deformation is most easily observable - in some cases reaching around a 3:2 ratio of the largest to smallest axis.  The deformations to planetary shape swamp even some putative atmospheric heights in these cases.  For example, for a 1.5 $R_{e}$ planet at the fluid Roche limit around a M5V star with an effective temperature of 2800, the atmospheric scale height of a pure sodium atmosphere \citep{mig11} at the substellar point of the planet is about 85 times smaller than the distortion of the planetary axis in that direction.  This would suggest the gravitational potential gradient that exists for atmospheric flow from the substellar point to the terminators of these planets would appear to be significant and may potentially affect observational atmospheric retrieval.  However, understanding of the dynamics of atmospheres of very hot close in planets is still very basic and would need to be explored in greater depth to derive any conclusions.  The transit bias for a particular mean radius planet due to the smaller observed mid-transit cross section of the distorted planet is given in figure \ref{fig:difftransitdepth}.  The bias is shown for 1, 1.5, and 2 $R_{e}$ planets and results are close for the other similar sized planets.  Dash-dot and dotted lines are given for 100 and 20 ppm thresholds which correspond to the \textit{Kepler Telescope's} \citep{bor10} combined differential photometric precision over a 6.5 hour integration for a 15th and 12th magnitude star respectively.

For the smallest star tested, the M6 V star, the differences in transit depth can be very significant and indeed for 1.5 $R_{e}$ planets and larger over an order of magnitude greater than projected \textit{TESS's} photometric precision of 200 ppm in 1 hour on an I=10 star.  Even for a 1 $R_{e}$ planet the transit depth difference can approach about 0.1\%.  For the M5 V star the effects are somewhat smaller but still observationally significant for even the 200 ppm threshold for a 1 $R_{e}$ planet.  The larger M1 V star may have biases which are not quite observationally significant for 1 $R_{e}$ but which become so for 1.5 $R_{e}$ planets and larger (in fact the 2 $R_{e}$ values reflected by the dashed green line are greater over equivalent distances then a 1.5 $R_{e}$ planet around an M5 V star).  This suggests that for M dwarf stars such a bias needs to be considered when looking at close-in planets.  In fact, for large solid planets 2 $R_{e}$ and greater it appears such an effect may even be detectable around larger K class dwarf stars.  Unsurprisingly, the effect is small around the G and F class dwarf stars.  Also as expected, biases were smaller for planets with greater rigidity and larger for less rigid planets. The transit depth differences can then be calculated as the corresponding radius underestimates for the planets in each system as given in figure \ref{fig:radiusunder}.

Figure \ref{fig:radiusunder} reflects the fact that planets which are the most distorted in this particular configuration will yield the most biased radius measurements.  In the case of the planets around the M5 V star, underestimates all reach about 1\% by 0.005 AU and steeply increase to greater than 7\% as they approach the fluid Roche limit.  Around the larger M1 V star the underestimates reach around 1\% between .007 and 0.01 AU and range from about 2.5 to 5.5\% at the fluid Roche limits.  Underestimates are small for larger stars with values ranging from 0.5 to 1.5\% at the fluid Roche limit for the G class dwarf star.

The radius underestimates result in density overestimates for these planets.  For a 1\% radius underestimate the calculated value for the density will actually be a 3\% overestimate of the real value.  For cases around the M5 V star density overestimates then range from 3\% at 0.05 AU to about 20-25\% greater than the actual value for a 7-10\% radius underestimate at the fluid Roche limit.  The 2.5-5.5\% radius underestimates around the M1 V star correspond to estimated density values which are 7-15\% greater than the actual value.  Clearly, for some of these planets their aspherical shape can lead to significantly overestimated bulk density values.

\section{\textbf{Photometric Signatures of Asphericity}}
\subsection{In Transit Signatures}

\begin{figure}
  \begin{flushleft}
      \includegraphics[angle=90,scale=0.40]{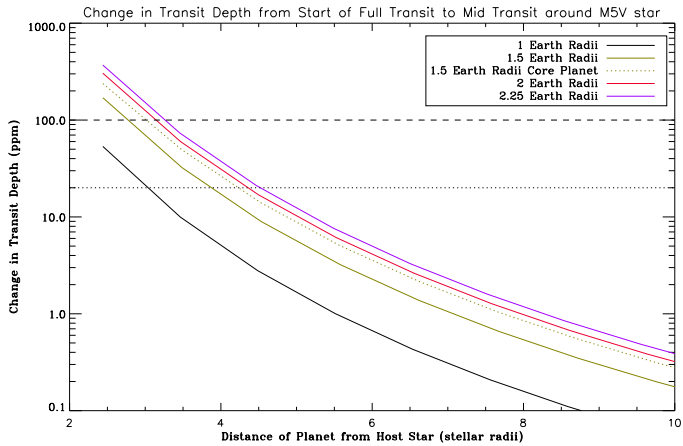}
      \caption{The change in transit depth from beginning of full transit to mid transit for different planets around an M5 V star.}
     \label{fig:signaturem5v}
  \end{flushleft}
\end{figure}

The synchronous rotation of these planets means that the projected area of a planet will appear to change as it travels around in orbit.  The maximum projected area will be at quadrature in the planet's orbit around its host star as the two projected axes will be the longest axis and the polar axis.  Conversely, at mid transit the projected area of the planet will be at a minimum as the visible axes will be the smallest axis and the polar axis.  This change in projected area has been previously recognized as a potential signature of planetary asphericity that is proportional during the transit portion of the light curve to $\sfrac{R_{s}}{a}$ \citep{rag09, car102} and has been generally assumed to be too minimal to use for system characterization.  In addition, there have been concerns that such a signal would in any event be degenerate with stellar limb darkening contributions.  However, the direct signal of an aspherical synchronous transiting planet should actually have the opposite effect of limb darkening by obscuring progressively less light through mid transit \citep[fig. 5]{lec11} – whereas stellar limb darkening effects result in a greater apparent stellar flux being blocked as the planet approaches mid transit.  For unknown limb darkening parameters this signal may be difficult to extract but as we will discuss in the next section there now exists the potential to disentangle such effects.  

Due to large bias values we saw for certain systems in the last section and the unique nature of a potential signal for an apsherical transiting planet we quantify the change in transit signal due to asphericity from the beginning of full transit to mid transit in the systems we examined.  The change in the projected area of the planet in this case (from the beginning of full to mid transit) is roughly proportional to one half the projected angle of the full orbit subtended by the star.  We calculate the total subtended angle of the transit as  

\begin{equation}\theta \approx \arcsin \dfrac{(R_{s}-R_{p})}{a}\end{equation}

This is an approximate expression because the angles are calculated using the mean radius of the planet when finding the start of full transit.  In reality at the beginning of full transit the planet will be rotated by the angle corresponding to the beginning of full transit and will have a slightly different b axis value - however the change in the area of the full planetary orbit that is subtended due to this effect is likely to be minimal (much smaller than change in total projected area of the planet as the planet transits).  From this angle we then calculate the total change in transit depth from beginning of full transit to mid transit due to the changing projected area of the planet caused by its asphericity.   We use the same equation as the one used to calculate the bias except the b axis is now rotated by the angle 90 - $\theta$ degrees for the projected area at the beginning of full transit and is compared to the mid transit projected area (with $\theta$ = 0 at mid transit) versus an idealized spherical case. Thus while the rotational contribution to the b axis remains the same at the beginning of full transit in this co-planar case the tidal contribution we add to it is now $S_{2}(P_{2}\cos (\dfrac{\pi}{2}-\psi))$. 

The results of these calculations show that the change in transit depth during the course of the transit is minimal for most of the stars with the exception of the M dwarfs.  Figure \ref{fig:signaturem5v} shows that close-in to smaller M dwarf stars the in-transit signature of a distorted planet may in fact be observationally detectable and significant.  In fact, most of the planets tested yield signatures which exceed the 20 ppm limit at distances $\sim 2$ times the fluid Roche limit.  Nearer to the fluid Roche limit the change in transit depth is much greater and even for the 1.5 Re case approaches 200 ppm.  These signatures are increased by nearly a factor of two for M6 V stars.  We also test whether such a direct signature can then also be used to constrain rigidity and find that near the fluid Roche limit the difference between the two planets meet or exceeds the 20 ppm threshold.  The green dotted line for a 1.5 $R_{e}$ Iron-Ni Core remnant diverges from the 1.5 $R_{e}$ planet with a rigidity value close to the Earth's as they approach the fluid Roche limit.

\begin{figure*}
  \begin{center}
      \includegraphics[scale=0.55]{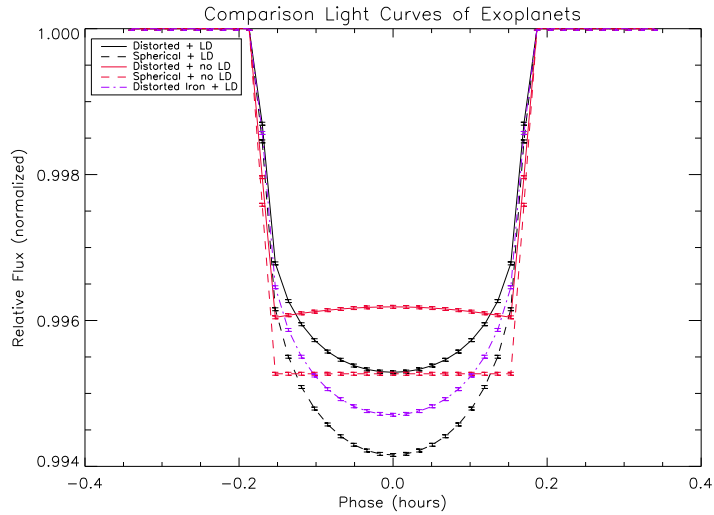}
      \caption{Light curves of a 1.5 $R_{e}$ planet transiting a M5V star at an orbital separation equal to the fluid Roche limit. Tidally and rotationally distorted planets with the empirically fit densities are given by the solid lines.  Spherical planets are given by the dashed lines.  The putative distorted Iron planet is given by the purple dash dot line.  Red lines correspond to cases with no stellar limb darkening while black lines correspond to cases with stellar limb darkening values produced by Exofast for the relevant stellar parameters. Error bars correspond to the 20 ppm uncertainty.}
     \label{fig:lightcurve}
  \end{center}
\end{figure*}

\subsection{Ellipsoidal Variations}

The complementary change in the projected area of these planets occurs in the out of transit portion of the light curve.  Since the area of the total orbit subtended by the in transit portion is typically small, most of the change in the projected area of the planet occurs when it is not transiting its host star.  As a result one can use the long to short axis ratio of the deformed planet (given in figure \ref{fig:axisratios}) to approximate the total change in the emitted flux contribution from the planet in the out of transit portion of the light curve.  Given a favorable planet to star contrast ratio in the infrared such changes may result in detectable ellipsoidal variations in the planet.  Such a variation would be immune to stellar limb darkening and may be an effective pathway to constraining the asphericity and subsequently bulk properties of the planet.  There may even be the potential to observe ellipsoidal variations in the star in order to constrain planetary bulk properties.  Using the expression given in equation 2 of \citet{fai11} a 1.5 $R_{e}$ planet at its fluid Roche limit around a M5V star may be expected to produce ellipsoidal variations in the star of $\sim 10$ ppm assuming a stellar limb darkening coefficient of 1 from \citet{cla98} and a stellar gravity darkening coefficient of 0.6.

\section{Discussion and Analysis}

From the results in the previous sections it is clear that for certain planet-star systems the tidal and rotational distortions to a solid planet's shape can result in underestimates of those planet's radii and subsequent values based off the inferred volume.   Additionally, in the case of planets that are close in to smaller M dwarfs the distorted shapes of the planets may actually produce an observable signature of the shape in the transit itself.  Statistical occurrence rates for such planets are currently incomplete due to a lack of data for mid to late M dwarfs \citep{dre13}.  However solid planets interior to their fluid Roche limit, planets with orbital separations with similar a/$R_{stellar}$ values and planets close to the fluid Roche limit for other stellar classes have all been found (in addition Dressing and Charbonneau find occurrence rates approach .01 for planets smaller than 1.4 $R_{e}$ on sub 1.2 day orbits). The variation of rigidity of a planet may produce a small but detectable signal in the cases that were tested as one gets very close to the fluid Roche limit, and again it is important to remember planets have also been detected interior to the fluid Roche limit (the inner distance bound).  Merely the constraining of tidal bulge amplitude along with Roche limit considerations may put meaningful limits on interior structure.  The ability to directly constrain the shape of a planet would provide clues towards tidal theory, the orbital configuration of the system and bulk properties of the planet. 

Our group's preliminary work (see figure \ref{fig:lightcurve}) on the nature of such a signature suggests that such an effect can make the transit portion of a light curve more box-like as a planet's decreasing projected area towards mid transit in effect compensates for some of the limb darkening effects - an effect that has already been noted in the case of WASP-12b \citep{cow12}.  Limb darkening effects are several factors greater than the effect of asphericity in figure \ref{fig:lightcurve} and as a result the distinctive W shape of the transit produced by asphericity is not visible and instead manifests itself as what may be interpreted as merely different stellar limb darkening parameter fits.  Specific studies of distorted planets in synchronous orbit may be able to use limb darkening studies of similar host stars \citep{mul13} to resolve the effect of the distortion, though care needs to be taken when fixing the limb darkening parameters \citep{csi13}.  Close-in planets may also typically be in multi planet systems with other planets farther out \citep{san14} which may enable comparative studies of the planets' transits to help glean shape effects from an ultra short period planet.  Limb darkening degeneracies may also be broken by obtaining light curves in multiple wavelengths using the \textit{Spitzer Space Telescope} \citep{wer04} and in the future, the \textit{James Webb Space Telescope} \citep{gar06}.  Finally, additional thermal and reflected phase variation effects of an aspherical planet \citep{bud11} may also be helpful in corroborating the in-transit signature of a planet's shape, particularly given that these extremely hot planets may lack degeneracy inducing atmospheres.  While ellipsoidal variations in the star which are induced by the planet's tidal effect may complicate this \citep{loe03} they may be able to be estimated.

The assumptions made to calculate the specific biases and signatures are not likely to hold for many systems.  However, the general physical principles will apply across a number of different planet-star systems and should be applicable when considering the more general question of determining the spin state of a planet.  Given that the equilibrium tide can produce potentially observable signatures for planets around some stars even when considering rotational deformation, such a signature may produce idiosyncratic signatures for planets in particular resonances or even more general orbital configurations.  This needs to be explored for simple cases including those that examine gas giants and then also for more realistic cases that vary orbital parameters.  Indeed, there have been dozens of Hot Jupiters found interior to 2.5 Roche radii of their stars \citep[fig. 4]{gil14}.  Finally, for solid planets it is clear that planetary systems around M dwarfs may be particularly amenable to exploration of the orbital state and bulk characteristics of a planet due to the compact nature of allowed systems, the tendency for such stars to host smaller planets and the simple fact that any planetary effect will have a much larger observational signature around these smaller stars.  This should encourage investigations of M dwarfs in addition to the current work of the MEarth survey \citep{nut08} and \textit{K2 mission} \citep{how14} and the planned survey of close and bright M dwarfs by \textit{TESS}.

\FloatBarrier

\label{lastpage}

\end{document}